# PARAMETER IDENTIFICATION OF PRESSURE SENSORS BY STATIC AND DYNAMIC MEASUREMENTS

*Steffen Michael[1], Steffen Kurth[2], Jens Klattenhoff[3], Holm Geissler[4], Siegfried Hering[5]*

[1]Melexis GmbH, Erfurt, Germany
[2]Fraunhofer Institute for Reliability and Microintegration, Chemnitz, Germany
[3]Polytec GmbH, Waldbronn, Germany
[4]SUSS MicroTec Testsystems GmbH, Sacka, Germany
[5]X-FAB Seminconductor Foundries AG, Erfurt, Germany

## ABSTRACT

Fast identification methods of pressure sensors are investigated. With regard to a complete accurate sensor parameter identification two different measurement methods are combined. The approach consists on one hand in performing static measurements – an applied pressure results in a membrane deformation measured interferometrically and the corresponding output voltage. On the other hand optical measurements of the modal responses of the sensor membranes are performed. This information is used in an inverse identification algorithm to identify geometrical and material parameters based on a FE model. The number of parameters to be identified is thereby generally limited only by the number of measurable modal frequencies. A quantitative evaluation of the identification results permits furthermore the classification of processing errors like etching errors.

Algorithms and identification results for membrane thickness, intrinsic stress and output voltage will be discussed in this contribution on the basis of the parameter identification of relative pressure sensors.

## 1. INTRODUCTION

The development of the two criteria costs and reliability is essential for the further growth of the MEMS market. Efficient test procedures can reduce costs significantly by the detection of faulty sensors before the subsequent packaging and assembly steps.

Up to now there are very few publications concerning wafer level test methods and suitable instrumentation [1, 2]. The presented indirect parameter identification by modal frequencies is taken up for the identification of pressure sensors. An high precision vibrometer measures optically the electrostatic excited out-of-plane vibrations. By means of a FE model the sensor parameters are extracted by the measured frequency response. The vibrations reflect the material and geometrical sensor parameters. Hence all sensor parameters can be identified generally by this indirect approach. The restriction is given by the accuracy to be obtained - the identification accuracy depends on the sensitivity of the modal frequencies versus the investigated sensor parameters. In case of the relative pressure sensor the identification accuracy of the membrane thickness is better than 2%.

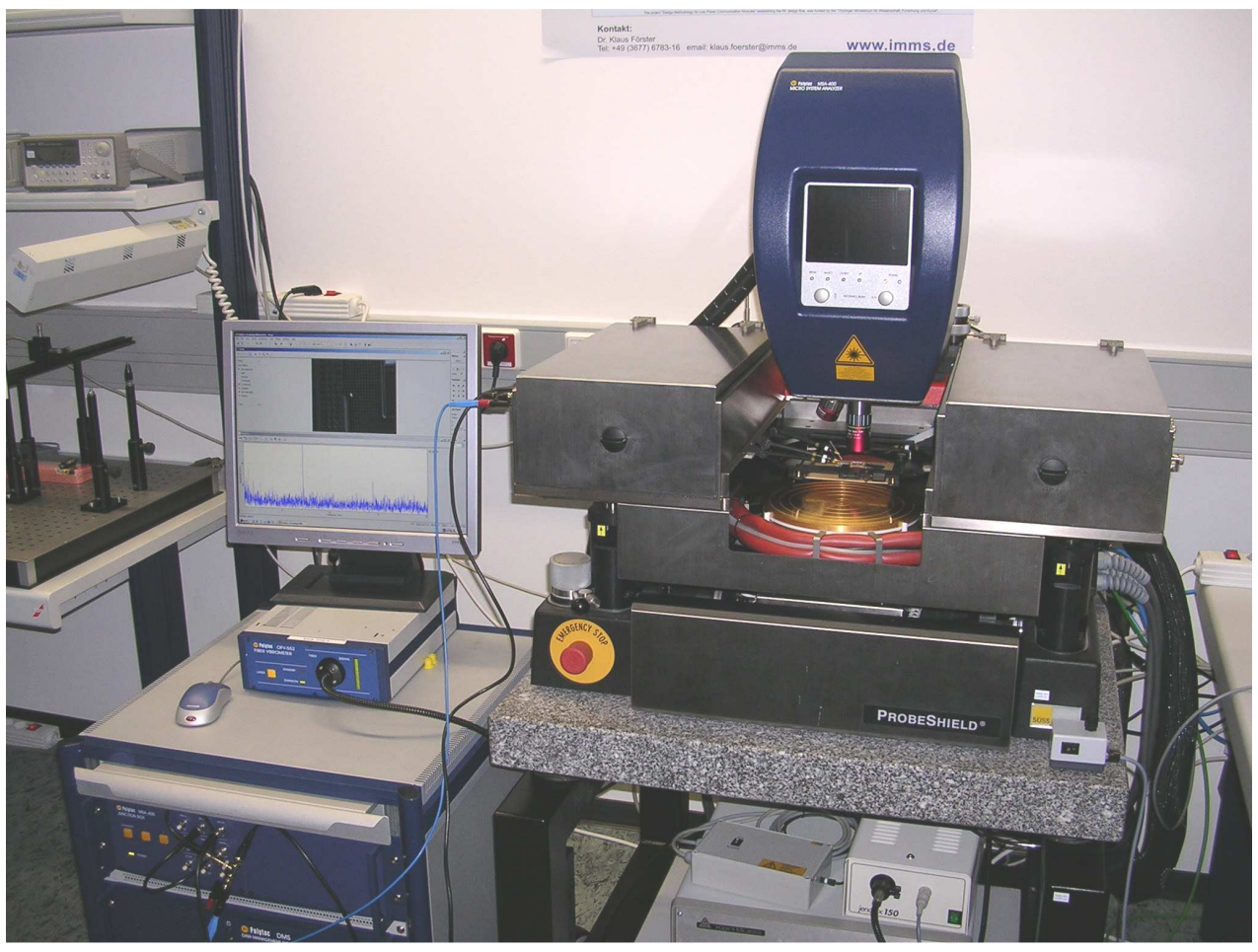

**Figure 1: Measurement setup**

The identification approach is well suited for the identification of whole wafers. The time consuming FE simulation has to be done only once for each sensor type. Furthermore the algorithm is based on the frequency response respectively the modal frequency values where a small number of measurement points (2-3) in the wafer test mode is sufficient. Due to a much longer measurement time resonant mode shapes are disregarded in the identification procedure.

Regular membranes are homogenous. Defects like etching errors cause an inhomogeneity. Such inhomogeneities will be detected by a quantitative evaluation of the identification results.

The output voltage respectively pressure sensitivity cannot be obtained directly by dynamic measurements. The static measurement yields the pressure sensitivity. The correlating data of both measurement methods permits an efficient sensor model improvement by adjusting e.g. piezoresistor parameters.

## 2. MEASUREMENT HARDWARE SETUP

The measurement setup is based on a semiautomatic Suss probe station and an optical micromotion analyzer MSA400 by Polytec [3]. Both systems are coupled with regard to an automated measurement of whole wafers.

The static measurements are done with the Suss Pressure Probe Module. An impact pressure is applied by a nozzle from the topside without touching of the die.

The out-of-plane vibrations with respect to the dynamic measurements are performed by an electrostatic excitation unit. Conventional excitation methods like ultrasound or piezoelectric stimulation cannot be used due to the expected modal frequencies in the range of MHz. The specified setup shown in Fig.1 enables the excitation and measurement of out-of-plane vibrations until 2.5MHz at wafer level.

### 2.1. Electrostatic excitation

The electrostatic excitation is realized by connecting an electrostatic probe to a high voltage (up to 400 volt) which is placed above the membrane surface. The gap between membrane and electrode should be small enough to yield a good SNR for an unique modal frequency peak detection. This is the only requirement because only the frequency but not the amplitude is used for the identification. Depending on the sensor type respectively membrane stiffness the gap size varies between 30µm and 100µm.

The electrostatic force depends on the capacitance between the electrodes, the displacement between the membrane surface and the electrode as well as the applied voltage. Hence any MEMS resonator is excited independent from the material it consists of.

Measurements are done with grinded probe needles and intertwined finger electrodes which have been fabricated on glass substrate. Indium tin oxide (ITO) has been used as electrode material and patterned to electrodes by photolithography and etching. Such electrodes are well suited for the excitation due to their transparency for the laser and the observation by microscope. Furthermore a large capacitance can be obtained due to the extensive electrode form.

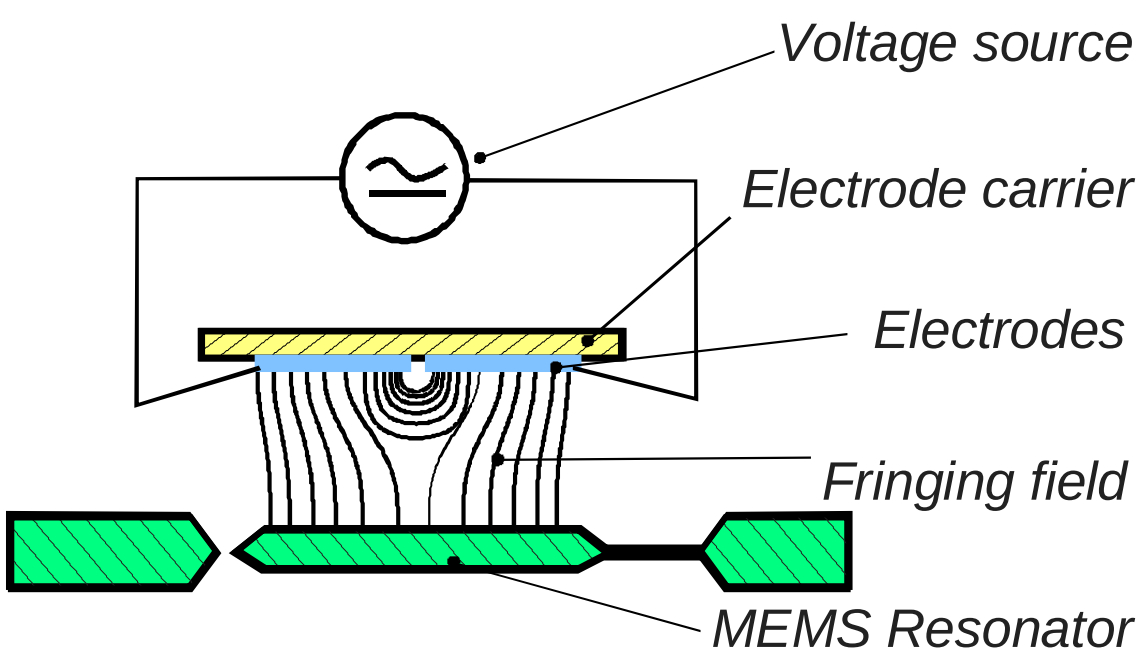

**Figure 2: Cross sectional view of fringing field electrode**

### 2.2. Pressure Probe Module

The pressure for doing static measurements is applied by a nozzle as a part of the Pressure Probe Module by Suss. By specifying a set pressure ($p_s$) at the base of the module's nozzle an air stream with the desired effective pressure ($p_w$) can be applied to the sensor under test. Via calibration the effective pressure is kept constant troughout the test cycle, irrespective of height differences on the wafer. The maximum pressure can be realized by the module is 7bar.

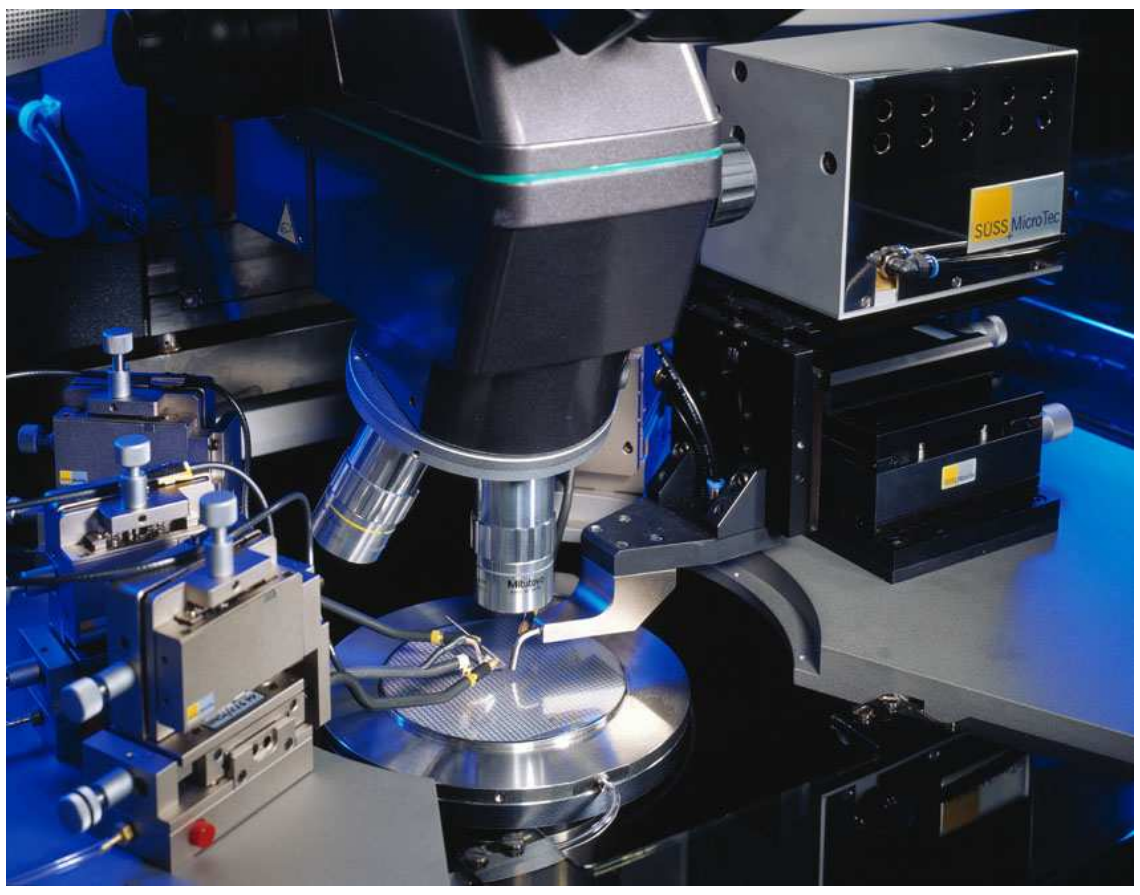

**Figure 3: Pressure Probe Module**

The pressure nozzle has to be adapted for each sensor type. With regard to an high accuracy the size of the pressure nozzle and the membrane have to match each other. Furthermore a minimum distance of 150µm between contact pads and the membrane has to be guaranteed to place the nozzle above the membrane surface.

The developed pressure module permits the simultaneous measurement of the bridge voltage and the membrane displacement in z-direction by an integrated glass window at the top of the nozzle.

### 2.3. Topographic and out-of-plane velocity measurements

The optical measurements are performed by the MSA400 Micro System Analyzer. In the measurement head of the Micro System Analyzer are integrated amongst other things a laser-Doppler vibrometer for fast and broadband out-of-plane dynamics as well as a white light interferometer for high resolution topography. By shifting an interference objective with nanometer precision with respect to the sample, a high resolution X-Y-Z mapping is generated. The objective focuses the interference pattern on to the camera.

For the out-of-plane measurements the laser beam of the vibrometer is scanned automatically over a user defined grid at the surface of the MEMS device. For each scan point an velocity measurement is performed by using the Doppler effect within some milliseconds. An internal generator controls the excitation of the pressure sensor by the electrostatic excitation unit and synchronizes the phase between the measurement points.

## 3. IDENTIFICATION ALGORITHM

The identification system based on dynamic measurements is subdivided into the modules measurement unit, simulation unit and the real identification module with the submodules peak detection, polynomial approximation and optimization.

The automated identification process can be controlled by the parameters

- Polynomial approximation accuracy,
- Sensor parameter range and
- Maximal *Estimated Identification Error (EIE)*.

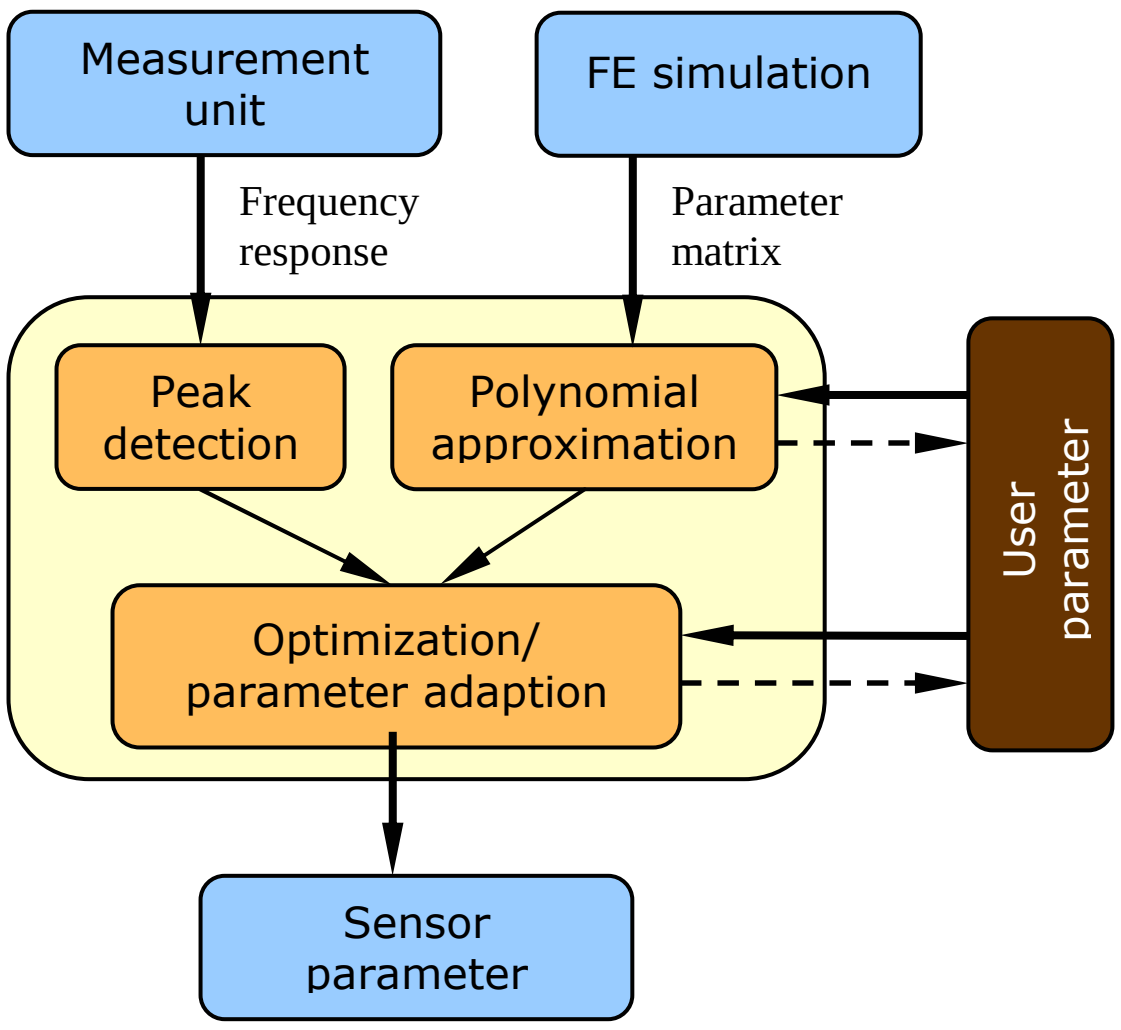

**Figure 4: Identification structure**

The term EIE is introduced with regard to a quantitative evaluation of the identification results. Considering the presented example of a relative pressure sensor where the two parameters membrane thickness and stress of the passivation layer have to be identified. Four modal frequencies are chosen for the identification. Hence six combinations of modal frequencies exist for doing the optimization. For each of the frequency combinations a parameter set $p_i$ is obtained. Then EIE is defined as

$$\boldsymbol{EIE} = \max(p_i) - \min(p_i)$$

and the normed $EIE_N$ correspondingly

$$\boldsymbol{EIE_n} = \frac{\max(p_i) - \min(p_i)}{\text{mean}(p_i)}$$

The EIE reflects the measurement and modeling errors and can be considered as a quantity of the identification accuracy. Typically EIE values depend on the number of identified parameters and the sensor type respectively the corresponding FE model.

Two different modes exist for the identification by modal frequencies – the characterization and the wafer test mode. The identification of a new sensor type starts within the characterization mode. A fine grid of measurement points permits an unique assignment of frequency peaks to the corresponding mode shapes. Suitable modal frequencies are chosen for the identification. Furthermore a maximal $EIE_N$ is defined as limit for valid dies based on the measurements of some dozen devices. Devices with a bigger EIE have defects not implemented in the FE model like etching errors and are declared as faulty dies. The identification of unknown but constant model parameters like stress is done within the characterization mode too.

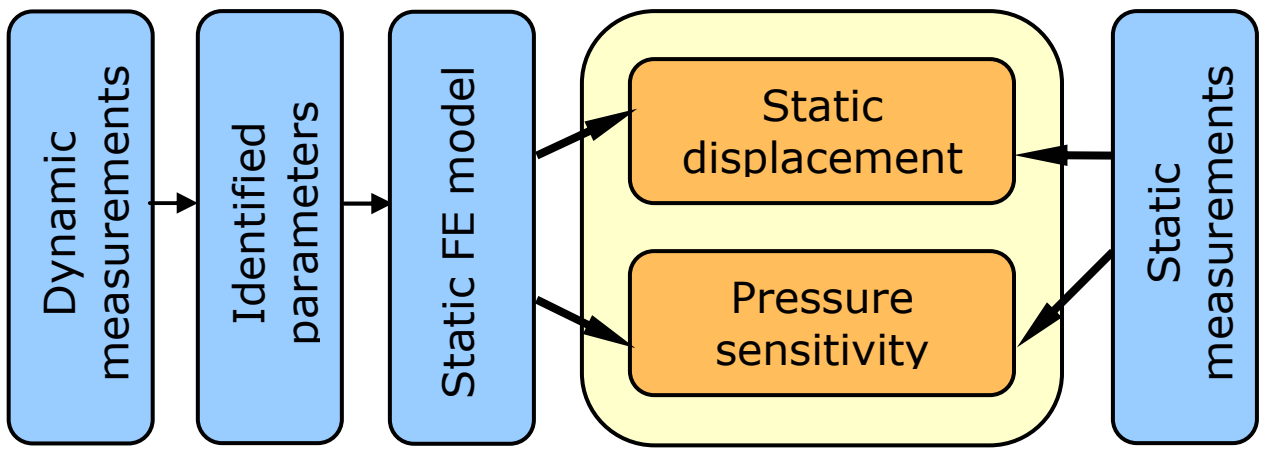

**Figure 5: Correlation of dynamic and static measurements**

Based on the results of the characterization mode measurements of full wafers are done in the wafer test mode. With regard to a minimal measurement time 2-3 measurement points are chosen. In the current configuration the measurement time per die is about 2-3 seconds. An improvement of the communication between prober and measurement system should reduce the measurement time to 1 second.

The output voltage respectively pressure sensitivity is not reflected directly in the measured modal frequencies. Like shown in Fig. 5 the pressure sensitivity is deduced by the identified sensor parameters by means of a static

FE model on the one hand. On the other hand the direct measurement of the static parameters permits the correlation of both measurement methods.

### 3.1. FE modeling and simulation

A common pressure sensor FE model implemented in *Ansys* is the base for different simulations – a modal, a static and a harmonic analysis.

The modal analysis is a nonlinear prestressed one due to the stress in the passivation layer. In case of the relative pressure sensor the modal analysis is performed with the membrane thickness and the stress as varying parameters. The parameter matrix obtained by the simulation is the base of the identification procedure.

The static analysis is done with regard to the membrane displacement and the stress distribution as well as the correspondingly piezo resistor changes.

An additional harmonic analysis is done in order to investigate the influence of the lateral electrode position to the modal frequency value. Due to the small damping the modal frequencies can be considered independently from the lateral electrode position which was confirmed by measurements.

### 3.2. Polynomial approximation

The parameter matrix given by the FE simulations is polynomially approximated with regard to a fast and accurate parameter identification by the least square method.

The polynomial approximation results usually in polynomials of higher orders. In case of an approximation of frequencies versus sensor parameters the polynomials of higher order will cause a time consuming nonlinear optimization. The inverse approximation – parameters versus frequencies – avoids a nonlinear optimization. In case of the presented two-dimensional problem the automated approximation results in a set of polynoms of $2^{nd}$ degree for each of the six frequency combinations (four modal frequencies are used)

$$z_i = c_{i,10} + c_{i,11} f_{i,1} + c_{i,12} f_{i,2} + c_{i,13} f_{i,1} f_{i,2} + c_{i,14} f_{i,1}^2 + c_{i,15} f_{i,2}^2 + c_{i,16} f_{i,1}^2 f_{i,2} + c_{i,17} f_{i,1} f_{i,2}^2 + c_{i,18} f_{i,1}^2 f_{i,2}^2$$

$$s_i = c_{i,20} + c_{i,21} f_{i,1} + c_{i,22} f_{i,2} + c_{i,23} f_{i,1} f_{i,2} + c_{i,24} f_{i,1}^2 + c_{i,25} f_{i,2}^2 + c_{i,26} f_{i,1}^2 f_{i,2} + c_{i,27} f_{i,1} f_{i,2}^2 + c_{i,28} f_{i,1}^2 f_{i,2}^2$$

with $z_i$ as membrane thickness and $s_i$ as stress of the frequency set i.

Based on an user defined accuracy (default value 0.1%) the degree of the polynomial is selected by the approximation module. A warning is generated by the module if the required accuracy is not reached or an oscillation occurs between the calculated reference points.

In such a case a recalculation of the parameter matrix obtained by FE simulation with closer reference points is indicated.

### 3.3. Peak picking

The measurement system provides a frequency response of the membrane structures. Two different peak picking algorithms were investigated with respect to speed and robustness – a fitting by a nonlinear least square algorithm with Gaussian or Lorentzian functions and a conventional local maximum search algorithm. Due to a faster peak detection the local maximum search algorithm is implemented in the submodule.

### 3.4. Optimization and parameter adaption

Identified peaks do not correspond necessarily with modal frequencies due to stochastically occurring noisy peaks. The unique assignment of modal frequencies to frequency peaks is done within the linear optimization where the minimization of the EIE is the objective function.

The submodule yields the sensor parameter and the corresponding EIE values. Furthermore a failure bit is initialized if the parameter or EIE limits are exceed.

## 4. APPLICATION EXAMPLE – RELATIVE PRESSURE SENSOR

Measurements are done at relative pressure sensors where the membrane is processed by KOH etching. For the identification 200 stochastically chosen dies are selected with a membrane size of 1300µmx1300µm. The membrane is covered by a passivation layer with several sublayers.

### 4.1. Identification by dynamic measurements

Two types of parameters to be identified can be distinguished. A constant but unknown FE model parameter is the stress in the passivation layer which is identified within the characterization mode. On the other hand there are parameters which vary due to the fabrication process like the membrane thickness. After the measurements of some test chips it seemed that lateral membrane dimensions have to be considered too by the identification process [4]. The measurements of 150 dies did not approve that so the membrane thickness is the unique process varying parameter to be identified.

Depending on the number of measurement points up to 10 modal frequencies can be measured in the frequency range until 1MHz. With regard to a minimum number of measurement points in the wafer test mode the first three modal frequencies are used for the identification.

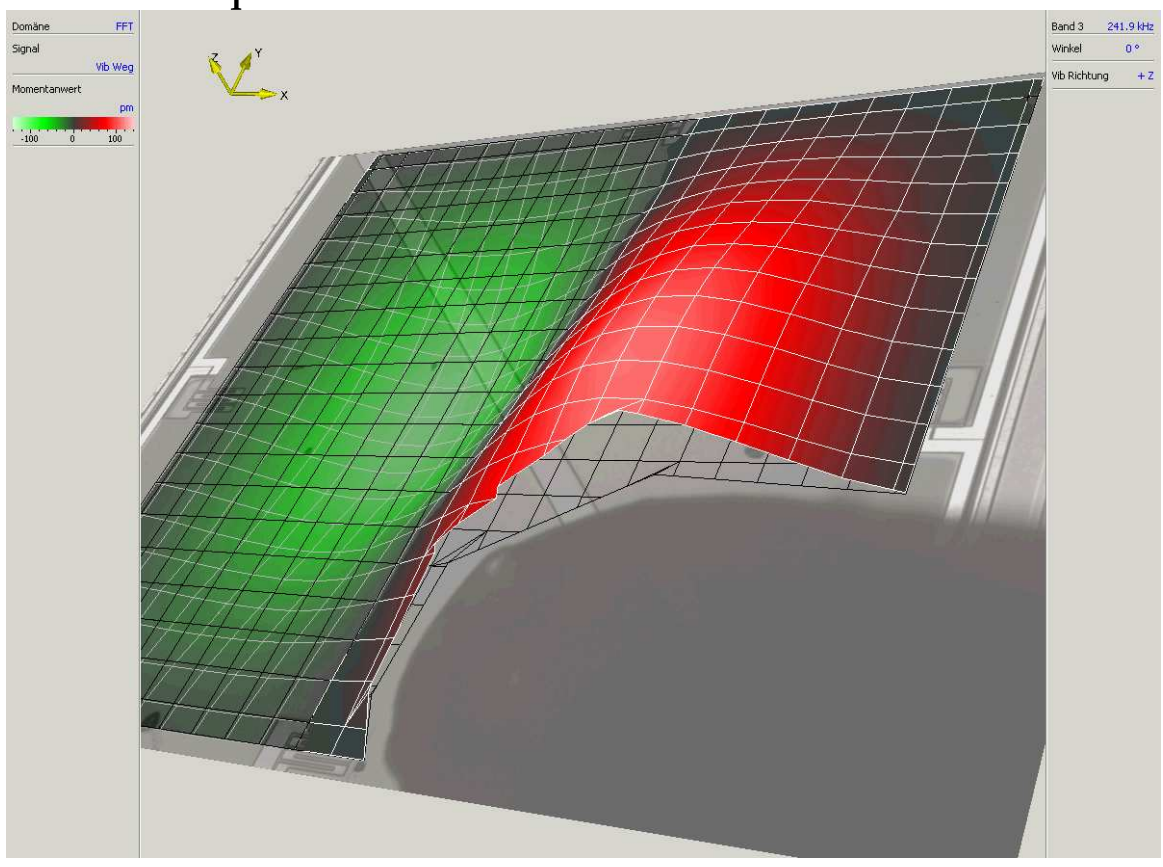

**Figure 7: Measured shape of 2nd modal frequency (characterization mode)**

Fig. 7 shows the identification results of the membrane thickness including a passivation layer of 4µm. The corresponding EIE of regular dies is less than 0.25µm - within the characterization mode the EIE limit for faulty dies is defined at 0.25µm. Hence the relative accuracy is about 2%. The EIE reflects beside the measurement errors model simplifications too. The FE model handles the passivation layer as a constant one, parameter changes due to technology reasons are neglected. This model simplification causes a relative large $EIE_N$ due to relative large ratio passivation versus membrane thickness. In opposite to relative pressure sensors the passivation versus membrane thickness ratio of absolute pressure sensors is much lower (1µm/19µm) which yields to an $EIE_N$ of 0.5% (0.1µm absolute).

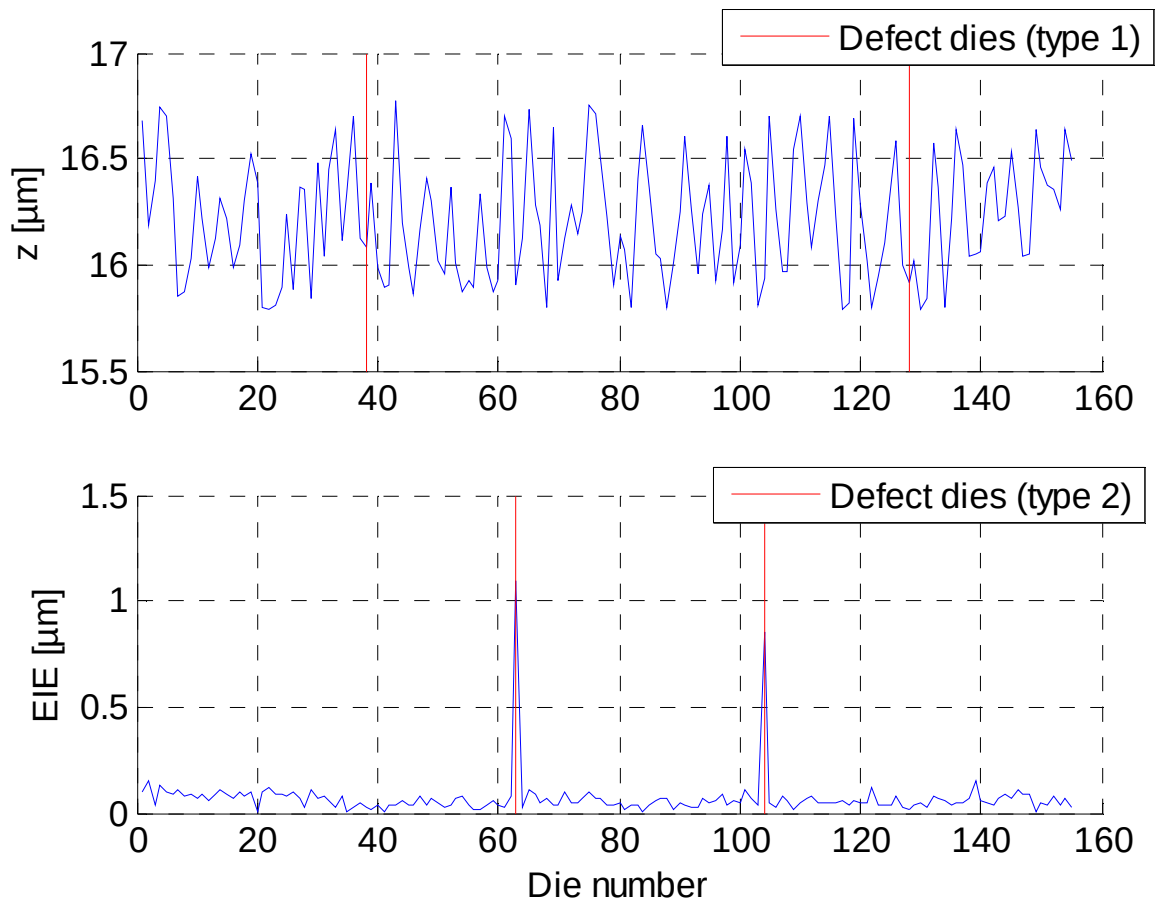

**Figure 8: Identified membrane thickness and corresponding EIE value**

The defect membranes identified by the system can be classified in two categories. Dies without a membrane like edge dies are easy to detect by the missing frequency peaks (type 1). The second group of faulty dies has an unsymmetric membrane probably due to an etching error.

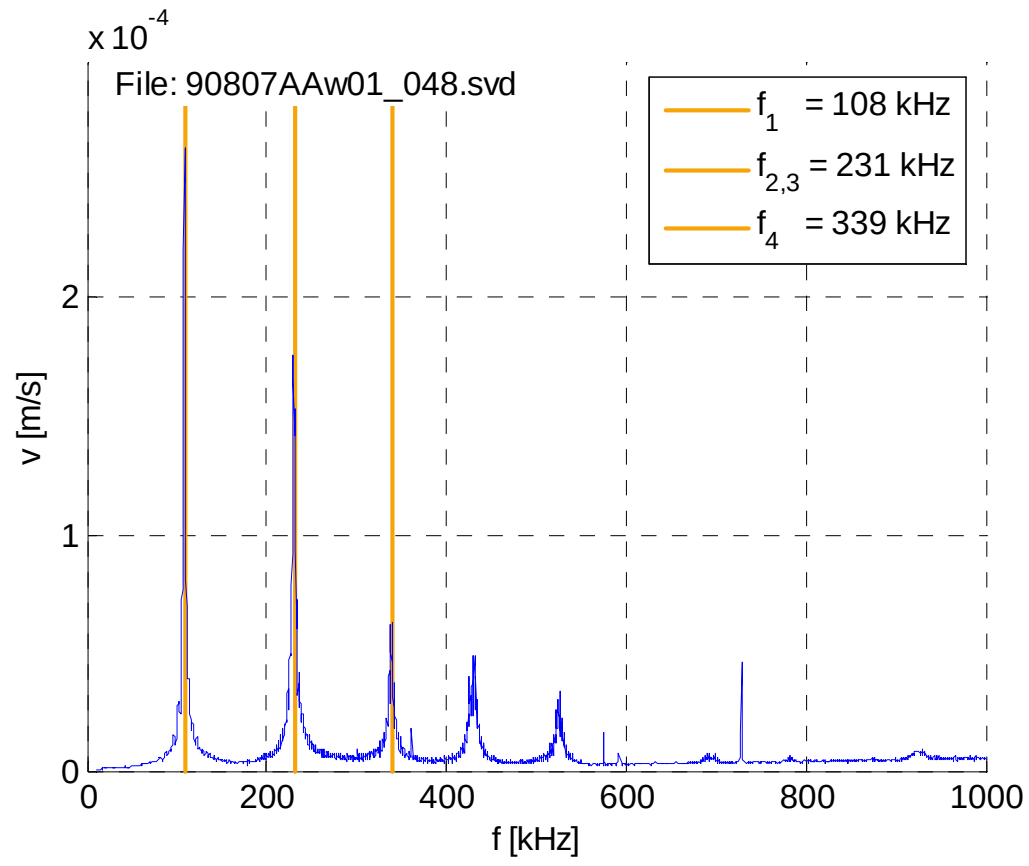

**Figure 6: Frequency response of a valid die**

Such an unsymmetry reflects in the splitting of the 2nd and the 3rd modal frequency, whereas the 2nd and the 3rd frequency of a quadratic membrane vary only in the phase but not the frequency value. The identification of such defect dies is realized via the maximal EIE criteria. For security reasons all defect identified membranes are checked by an additional mode shape test. The form of the mode shapes approved the classification as defect dies.

### 4.2. Identification by static measurements and data correlation

Static measurements are done with an applied pressure from 0 to 0.5 bar. The bridge output voltage and deformation are measured simultaneously which is possible due to the glass at the top of the pressure nozzle.

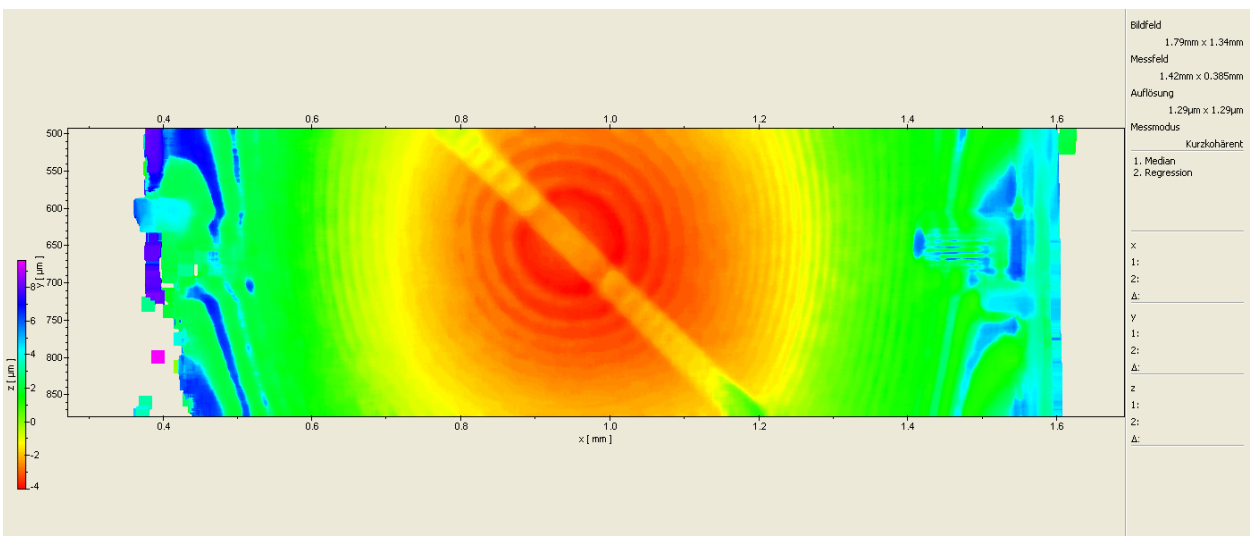

**Figure 9: Measured z displacement by applied pressure**

The measured static parameters are compared with the simulated one determined by the dynamically identified sensor parameters. In Fig. 10 a membrane thickness of 15.3 µm is identified for die #35. Based on the identified thickness the bridge output voltage is simulated by a static analysis which shows a good correlation between simulated and measured voltage after the adaptation of piezoresistor parameters like effective implantation depth.

The difference between measured and simulated voltage is less then 10% within all investigated 150 dies.

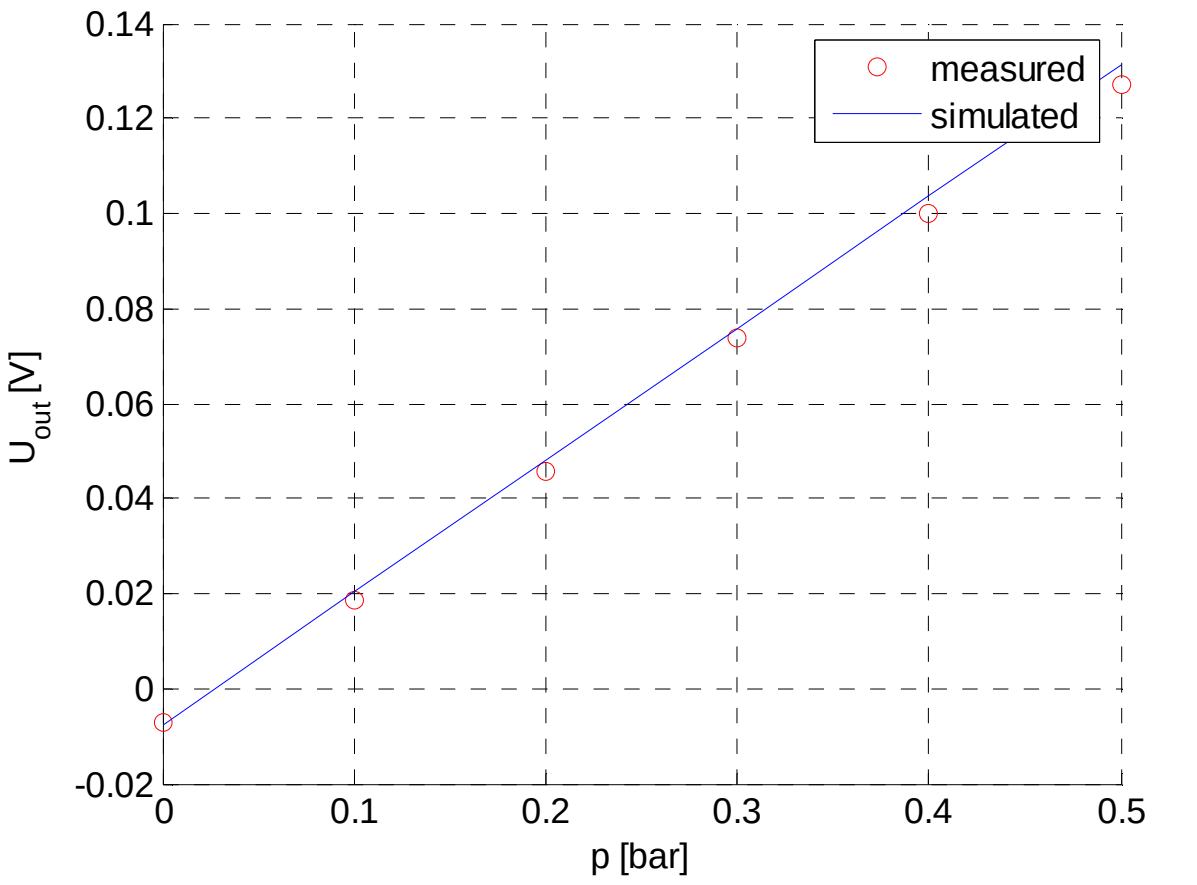

**Figure 10: Measured bridge output voltage versus simulated one by identified parameters (die #35)**

## 5. CONCLUSIONS

The measurements showed that both types of identification are well suited for the application in the production process due to their accuracy and speed. The dynamic measurements enable the identification of several geometric parameters whereas the static measurements provide only one but the interesting parameter sensitivity. The preferred identification method depends therefore from the sensor type and the interesting parameters.

## 6. ACKNOWLEDGEMENT

The authors acknowledge the financial support by the Federal Ministry of Education and Research Germany for the research project PARTEST.